\documentclass[epj]{svjour}
\usepackage{graphics}
\begin{document}
\title{High-spin structures of $^{86,87,88,89}$Y: a shell model interpretation}
\author{}
\author{P. C.~Srivastava$^1$ \thanks{e-mail: pcsrifph@iitr.ac.in}, Vikas Kumar$^1$ 
\and M. J. Ermamatov$^2$}
\institute{Department of Physics, Indian Institute of Technology, Roorkee 247 667, India
\and Institute of Nuclear Physics, Ulughbek, Tashkent 100214, Uzbekistan}

\date{\today}

\abstract{   
In this work nuclear structure properties of $^{86,87,88,89}$Y isotopes have been
investigated using large-scale shell-model calculations within the
full $f_{5/2}pg_{9/2}$ model space.  The calculations have been performed with
JUN45 and jj44b effective interactions that have been proposed for use in the
$f_{5/2}$, $p_{3/2}$, $p_{1/2}$, $g_{9/2}$ model space for both protons and neutrons.
Reasonable agreement between experimental and calculated values are obtained. 
This work will add more information to previous study by projected shell model
[Eur. Phys. J. A 48, 138 (2012)] where full-fledged shell model calculations proposed for these nuclei.
}

\authorrunning{P.C. Srivastava}
\maketitle

\maketitle

\section {Introduction}
 \label{s_intro} 
 
The neutron rich nuclei with $Z$ = 28-40 recently  attracted much theoretical and  experimental affords 
\cite{hagen,kaneko,sorlin,holt,lenzi,steppenbeck,zamick,sri1,sri2,sri3}.
Many fascinating phenomena have been observed in this region. The nuclei Sr, Y and Zr  are close
to subshell closure, thus they are expected to exhibit single-particle characteristics.  The 
other important features in this region are existence of high-spin isomers and shape transitions 
as higher $j$ orbitals are occupied. Recently structure of Sr and Zr isotopes near and at the magic
number $N=50$ shell via $g$-factor and life-time measurements have been investigated. 
The Zr isotopes changing structure from $^{80}$Zr (super deformed) with high occupancy of $\pi g_{9/2}$
orbit to less deformed $^{90}$Zr with substantial $\pi g_{9/2}$ orbit occupancy.
Recent experimental study claim the evolution of collectivity and shape coexistence for Sr isotopes.

The high-spin states in $^{86}$Y using heavy-ion fusion-evaporation reactions have been studied
in \cite{ref86Y1}. The corresponding structures were interpreted by shell-model with truncation.
For $^{88}$Y the high-spin states up to an excitation energy of 8.6 MeV and spin and parity of 
19$^{(-)}$ reported in Ref. \cite{ref1.88Y}. In the another experiment high-spin states up to spin 
21$\hbar$ investigated through fusion-evaporation reaction $^{82}$Se ($^{11}$B, 5n) \cite{ref2.88Y}.
The excited states of $^{87}$Y up to 33/2$^{(-)}$ at $\sim$ 7 MeV with in-beam $\gamma$ ray spectroscopy
reported in \cite{ref3.87Y}.  In this work majority of the observed high-spin states explain on the basis
of $\nu$ = 3 and $\nu$ =5 configurations.
Previously this nucleus were experimentally studied in Refs. \cite{ref4.87Y,ref5.87Y,ref6.87Y}, and
excited states up to $\sim$ 4.6 MeV were observed.  The experimental results  up to  31/2$^+$ $\hbar$
for $^{89}$Y is reported  in \cite{ref7.89Y}
 by using in-beam $\gamma$-ray spectroscopy.  B. Cheal {\it et al.} used laser spectroscopy method
 to study isomeric states of yttrium isotopes \cite{ref8.Y}. In this work nuclear charge radii differences,
 magnetic dipole and electric quadrupole moments have been obtained.
Recently, theoretical results of positive - parity yrast bands of odd $^{79-89}$Y isotopes using projected shell model
(PSM)  reported in \cite{ref9.psm}.  In this work it is mentioned that the results of large-scale shell
model calculation in this mass region is limited due to involvement of $g_{9/2}$ orbital
which generate large configuration space. Thus results of modern shell-model calculations
are desire for these nuclei.

In the present paper, we reported systematic study of shell-model results for $^{86,87,88,89}$Y isotopes.
The main motivation of present study to explain recently available experimental data for
these isotopes. 

The result of present work is organized as follows: Model space and Hamiltonian is given in section 2. In section 3-5,
energy levels and transition probabilities, quadrupole and magnetic moments, occupancy are compared with the available
experimental data. Finally, concluding remarks are drawn in section 6.


\section{Model space and Hamiltonians}\label{discuss_SM}

The present shell model (SM) calculations have been carried out  with two recently available
effective SM interactions, JUN45 and jj44b, that have been proposed for the $1p_{3/2}$, $0f_{5/2}$, $1p_{1/2}$ and
$0g_{9/2}$ single-particle orbits. JUN45, which was recently developed by Honma et al \cite{Honma09}, is
a realistic interaction based on the Bonn-C potential fitting 400 experimental binding and
excitation energy data with mass numbers A = 63$-$96. Since the present model space is not
sufficient to describe collectivity in these regions, data have not been used while fitting
in the middle of the shell along the $N = Z$ line. For JUN45, the data mostly fitted to develop
this interaction closure to $N = 50$.  This interaction is not successful in explaining
data for Ni and Cu isotopes, possibly due to the missing $0f_{7/2}$ orbit in the present model space.
The jj44b interaction due to Brown et al \cite{brown} was developed by fitting 600 binding energies and
excitation energies with $Z = 28$-$30$ and $N = 48$-$50$. Instead of 45 as in JUN45, here 30 linear
combinations of good $J–T$ two-body matrix elements (TBME) varied, with the rms deviation of
about 250 keV from experiment. For the JUN45 interaction, the single-particle energies are
taken to be -9.8280, -8.7087, -7.8388 and -6.2617 MeV for the $p_{3/2}$, $f_{5/2}$, $p_{1/2}$ and $g_{9/2}$ orbits,
respectively. Similarly for the jj44b interaction, the single-particle energies are taken to be
-9.6566, -9.2859, -8.2695 and -5.8944 MeV for the $p_{3/2}$, $f_{5/2}$, $p_{1/2}$ and $g_{9/2}$ orbits, respectively.

All calculations in the present
paper are carried out at TLAPOA computational facility
at ICN, UNAM, Mexico
using the shell model
code \textsc{antoine}.\cite{Antoine} In case of $^{79}$Se for positive parity
maximal dimension is 59 791822. 



\begin{figure*}
\resizebox{0.90\textwidth}{!}{
\includegraphics{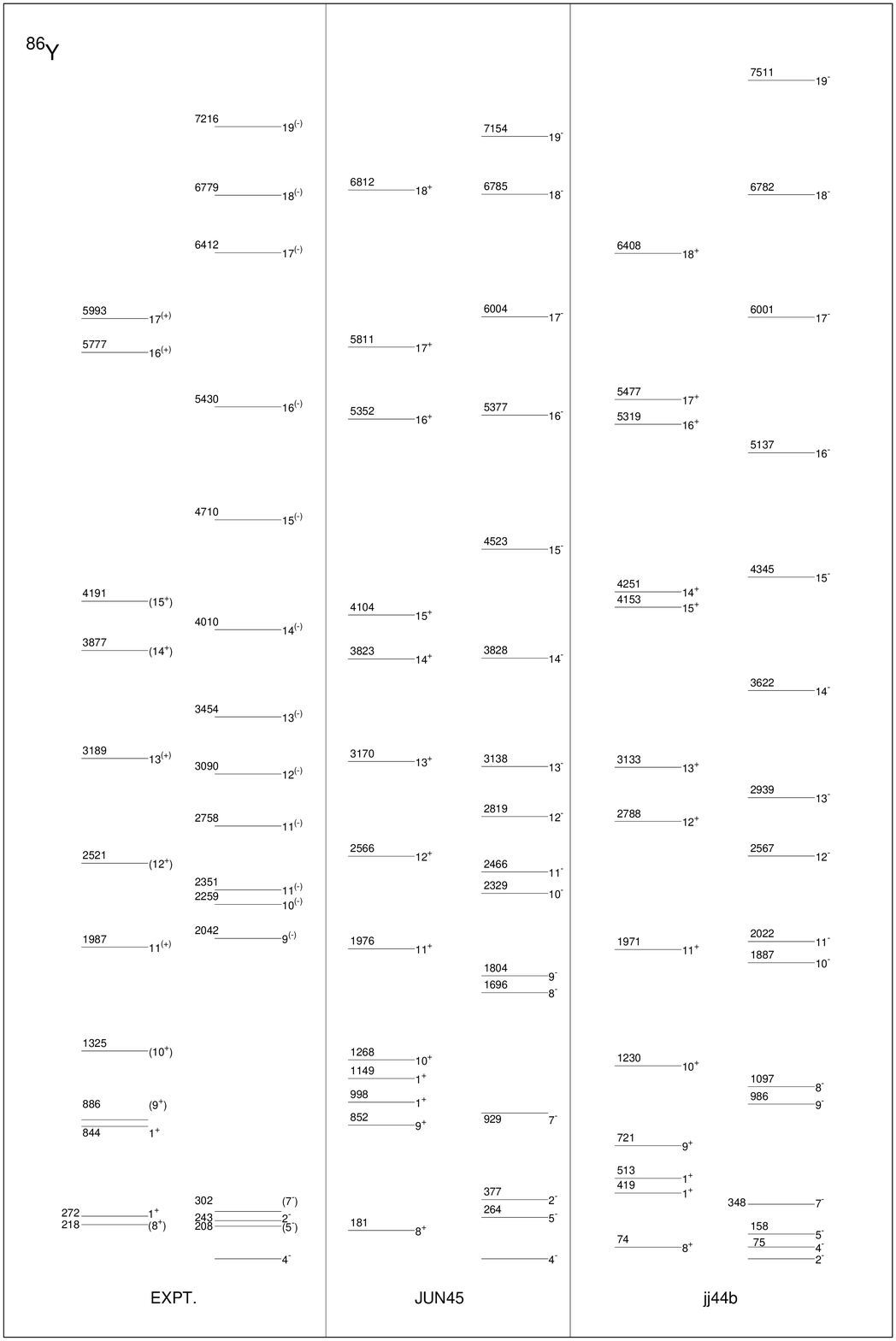} 
}
\caption{Comparison of shell-model results with experimental data for $^{86}$Y with different interactions.}
\label{fig1}
\end{figure*}
\begin{figure*}
\resizebox{0.90\textwidth}{!}{
\includegraphics{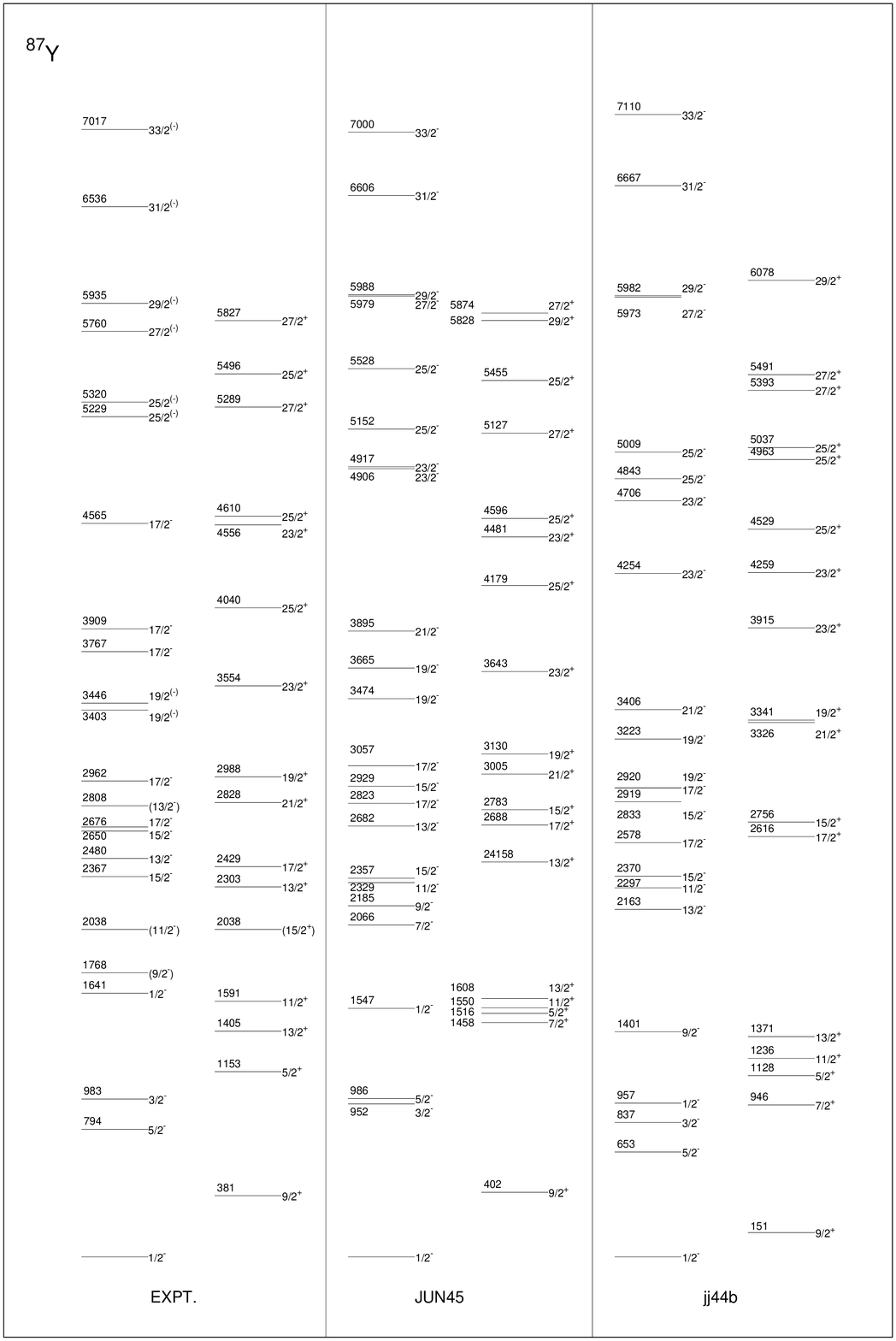} 
}
\caption{Comparison of shell-model results with experimental data for $^{87}$Y with different interactions.}
\label{fig2}
\end{figure*}

\begin{figure*}
\resizebox{0.90\textwidth}{!}{
\includegraphics{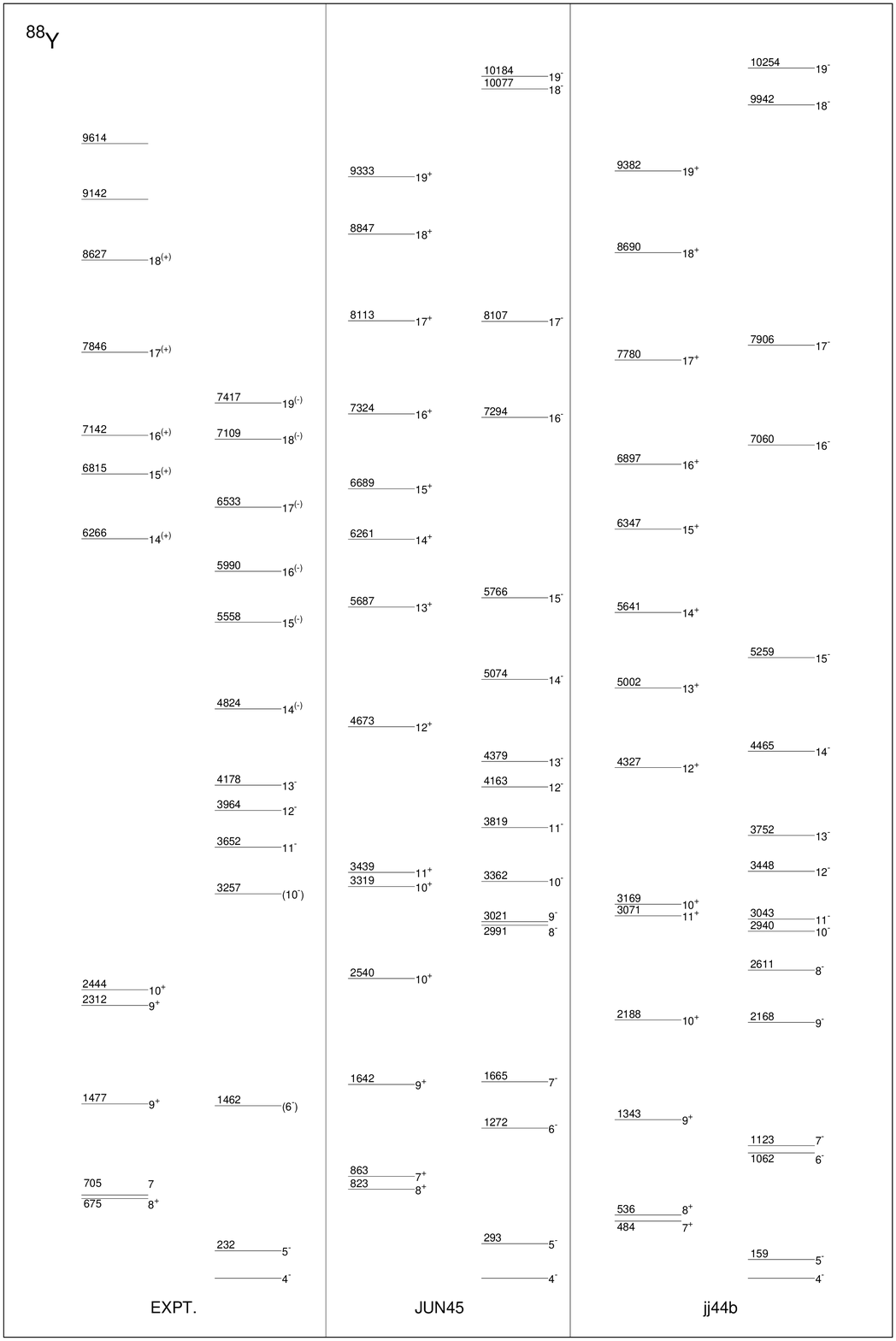} 
}
\caption{Comparison of shell-model results with experimental data for $^{88}$Y with different interactions.}
\label{fig3}
\end{figure*}

\begin{figure*}
\resizebox{0.90\textwidth}{!}{
\includegraphics{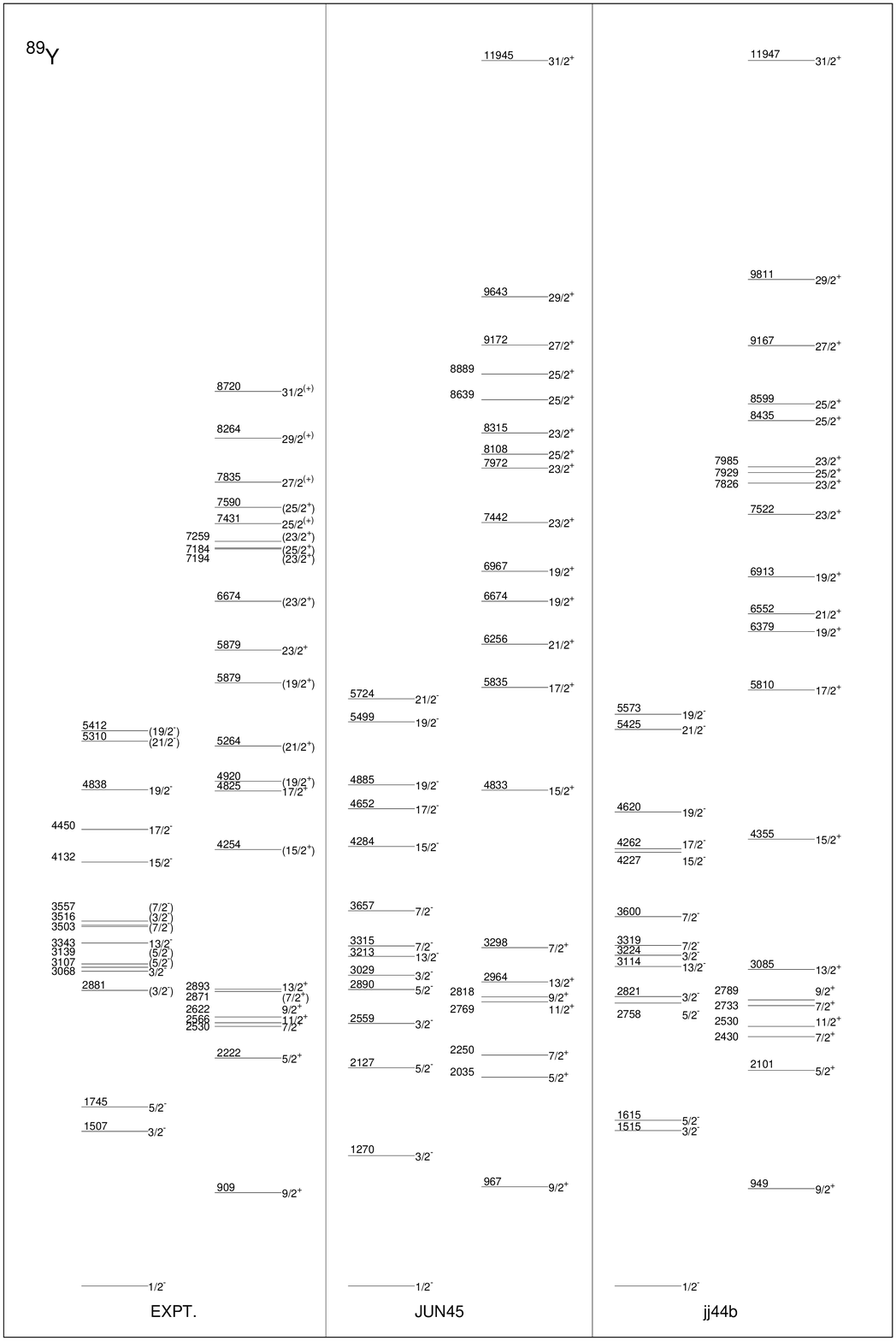} 
}
\caption{Comparison of shell-model results with experimental data for $^{89}$Y with different interactions.}
\label{fig4}
\end{figure*}
\section{ Result of Y isotopes }\label{discuss_odd}

The shell model results for $^{86-89}$Y  with two different interactions are shown in Figs. 1--4.

\subsection{ $^{86}$Y }\label{discuss_odd86}

Previously shell model calculation in $f_{5/2}pg_{9/2}$ space for this isotope with truncation
by allowing up to two particles excitation from $f_{5/2}$ and $p_{3/2}$ to $p_{1/2}$ and $g_{9/2}$
is reported in Ref. \cite{ref86Y1}. The signature splitting and magnetic rotation of 
$^{86}$Y using self-consistent tilted axis cranking calculations based on relativistic mean field
theory to investigate the dipole structures have been studied by Li {\it et al} \cite{ref86Y2}.
In present study we performed shell model calculation in $f_{5/2}pg_{9/2}$ space, this will add more 
information in the previous study \cite{ref86Y1} where truncated shell model results reported.

The JUN45 predicts the same sequence of the first four levels as in the
experiment, however $7^-$ level is much higher than in the experiment.
This level is better predicted by jj44b, but the other two levels are
lower than in the experiment. All
the calculated levels up to $17^-$ have lower values in both calculations than
in the experiment. Then the value of $18^-$ comes very close to the
experimental one. The $19^-$ is little bit lower in JUN45 calculation
and in jj44b the calculated value of this level becomes $\sim$300 keV
larger as compared to experimental one.

The $8^+$ level at 218 keV predicted to be 181 and
74 keV by JUN45 and jj44b calculations, respectively.  Then sequence
of two $1^+$ levels are the same in jj44b calculation. The value of
the first $1^+$ level in this calculation is lower than in the experiment
and that of second level is higher. In JUN45 both levels are located higher
than in the experiment. A very good agreement is given by both calculation
starting from $9^+$ until $15^+$. Then both calculations starts to differ
from the experiment. For the ground state $4^-$, the JUN45 interaction predicting 
$ \pi (p_{1/2}^1) \otimes \nu (g_{9/2}^{-3})$ ($\sim$ 30 \%)  configuration while jj44b as 
$ \pi (g_{9/2}^3) \otimes \nu (f_{5/2}^{-1})$ ($\sim$ 17 \%).

\subsection{ $^{87}$Y }\label{discuss_odd}
Both calculations predict correct ground state. 
In the jj44b calculation the sequence of next four negative parity levels are
the same as in the experiment, though the values of
this levels are lower than the experimental ones. The $13/2^-$ level is lower
than in the experiment while $11/2^-$ is higher than in the experiment, but the
difference in $15/2^-$ experimental and calculated values is  only 3 keV.
 The $23/2_2^-$
levels which appear in the calculation have not been measured in the
experiment. Then the values of next $25/2_2^-$ levels are $\sim$ 300 keV lower
than in the experiment. Last four $27/2^-$, $29/2^-$, $31/2^-$ and $33/2^-$ levels
are in good agreement with the experimental data. 

In JUN45 calculation
the pair of $3/2^-$, $5/2^-$ and $1/2^-$, $9/2^-$ levels are
interchanged as compared to the experiment. The $1/2_2^-$ level is
much closer to the experiment than in jj44b calculation.

Positive parity levels start from $9/2^+$ in the experiment and in
both calculations. In jj44b it starts from lower value than in the
experiment. The $7/2^+$ level appearing in both calculations has not
been measured in the experiment. Then $5/2^+$ and $13/2^+$
levels are lower than in the experiment in jj44b calculation while
they are higher in the JUN45 calculation. Better agreement with the
experiment in higher spins gives JUN45 calculation.
The structure of ground state  $1/2^-$ is a single-particle character
($ \pi (p_{1/2}^1) $). The JUN45 and jj44b interactions predicting
$\sim$ 41 \% and  $\sim$ 22 \% probability, respectively.

\subsection{ $^{88}$Y }\label{discuss_odd88}

The first three negative parity levels are in better agreement with the experiment
in JUN45 calculation, while the values of these levels are smaller in the jj44b
calculation. The $7^-$, $8^-$ and $9^-$ levels which appear in both calculation
have not been measured in the experiment. Then the levels up to the $15^-$ are
better predicted by JUN45 calculation. The other calculated higher spin levels
have larger values than in the experiment.

The sequence of the first two positive parity levels is the same as in the
experiment in JUN45 calculation and the values of these levels are higher
than in the experiment. In the jj44b calculation these levels are predicted
lower and $7^+$ and $8^+$ levels are interchanged with respect to the
experimental ones. The $9^+$ levels is located higher in JUN45, while
it is lower in jj44b calculation. The value closer to the experimental one is
predicted by JUN45 calculation for $10^+$ level. The $10^+_2$, $11^+$, $12^+$
and $13^+$  levels which appear in the calculation  have not been measured
in the experiment. The levels predicted by JUN45 from $16^+$ to $18^+$ have larger values as
compared to experimental ones, while $16^+$ and $17^+$ have smaller values than in the
experiment in jj44b calculation. The $19^+$ level in both calculations is
close to the experimental level at 9142 keV, to which in the experiment
spin has yet not been assigned.  
Both interactions predicting $ \pi (p_{1/2}^1) \otimes \nu (g_{9/2}^{-1})$ configuration
for ground state $4^-$ 
 with probability  $\sim$ 57 \%  (JUN45)
and $\sim$ 46 \%  (jj44b).

\subsection{ $^{89}$Y }\label{discuss_odd}
Both calculations predict $1/2^-$ ground state as in the experiment.
In the JUN45 calculation the spacing between $3/2^-$ and $5/2^-$ levels
are much higher then in the experiment. In the jj44b calculation these two
levels are closer to the experiment and little bit compressed as compared
to the experiment.  The $3/2_1^-$  and $3/2_2^-$ predicted better by jj44b calculation.
The $5/2_2^-$ level in the experiment predicted lower by JUN45 and even
more lower by jj44b. Both calculations gives small difference in the
sequence of the next experimental $13/2^-$, $(7/2)^-$, $(3/2)^-$ and $(7/2)^-$ levels,
though the location of $3/2^-$ level looks better in jj44b calculation.
The sequence of last five experimental levels are exactly the same as in
the experiment in jj44b calculation and the values of these levels are
closer to the experimental ones than in JUN45 calculation.
  
Agreement of positive parity levels with the experimental ones become
much improved in both calculation for the $^{89}$Y as compared to
$^{87}$Y  until $15/2^+$. Then both calculations predict higher values
than in the experiment. As we move from $^{87}$Y to $^{89}$Y the ground state $1/2^-$ is the same
with configuration i.e. $ \pi (p_{1/2}^1) $, now probability become 
$\sim$ 71 \%  (JUN45) and  $\sim$ 67 \%  (jj44b).


 \begin{table}
\begin{center}
\caption{$B(E2)$ reduced transition strength in W.u. Experimental values were taken from
  the NNDC database.}
\vspace{2mm}
\label{tab:table2}
\resizebox{!}{3.5cm}{
\begin{tabular}{ c  c  c  c  c  c } \hline 
\vspace{0.2cm} 
        & $I_1^\pi \rightarrow I_f^\pi$ & $E_\gamma$ & Exp. & JUN45 & jj44b  \\ \hline
$^{86}$Y&$4^- \rightarrow 2^-$ & 242.80  & 1.00 (8) & 2.12 & 0.08\\ 
        &$(8^+) \rightarrow (10^+)$ & 1107.08  & $>0.030$ & 11.67 & 18.06\\ 
        & $(10^+) \rightarrow (12^+)$ & 1195.88  & $>0.021$ & 10.74 & 22.04\\
\hline 
$^{88}$Y &$(8^+) \rightarrow (10^+)$ & 1769.35  & $>0.6233$ & 9.30 & 9.34\\ 
         &$(10^+) \rightarrow (12^+)$ & 1195.88  & $>0.021$ & 5.44 & 0.91\\ 
\hline                
$^{87}$Y &$1/2^- \rightarrow 5/2^-$ & 793.7 & $\geq 0.0078$ & 6.54 &  1.06\\ 
        &$13/2^+ \rightarrow 17/2^+$ & 1023.6 & $\geq 0.0022$ & 7.53 & 18.30\\
        &$17/2^+ \rightarrow 21/2^+$ &399 & 4.6 (3) & 3.99 & 5.00 \\ 
        &$21/2^+ \rightarrow (23/2^+)$ & 159.8 & $\geq 4.1$ & 7.51 & 9.90 \\ 
        &$21/2^+ \rightarrow (25/2^+)$ & 1782.4 & 4.9 (21) & 5.67 & 6.76\\ 
\hline       
$^{89}$Y&$9/2^+ \rightarrow 5/2^+$ & 1313.2 & 4.3 (13) &   2.99 &  3.22\\ 
        &$9/2^+ \rightarrow 13/2^+$ & 1984.1 & 4.3(8) & 7.98 & 8.50\\
        &$13/2^+ \rightarrow 17/2^+$ & 1931.9 & $<0.029$ & 0.98 & 6.69 \\ 
        &$21/2^+ \rightarrow (25/2^+)$ & 1920.2 & 4.4(18) & 5.94 & 3.26 \\ 
        &$25/2^{(+)} \rightarrow (29/2^+)$ & 832.9& $<8$ & 0.53 & 5.84 \\ 
        &$27/2^{(+)} \rightarrow (31/2^+)$ & 885.9 & $<19$ & 1.09 & 0.02  \\ 
        &$1/2^- \rightarrow 5/2^-$ & 1744.7 & 2.3 & 5.38 & 5.29   \\ 
        &$13/2^- \rightarrow 17/2^-$ & 1106.5 & 2.1(5) & 2.24 & 0.12 \\ 
        &$15/2^- \rightarrow 19/2^-$ & 706.3 & 2.2(10) & 1.55 & 3.09 \\ 
        &$17/2^- \rightarrow (21/2^-)$ & 860.1 & 0.57(15) & 0.93 & 2.27  \\ 
\hline
 \end{tabular}}
 \end{center}
\end{table}

\begin{figure*}
\resizebox{1.00\textwidth}{!}{
\includegraphics{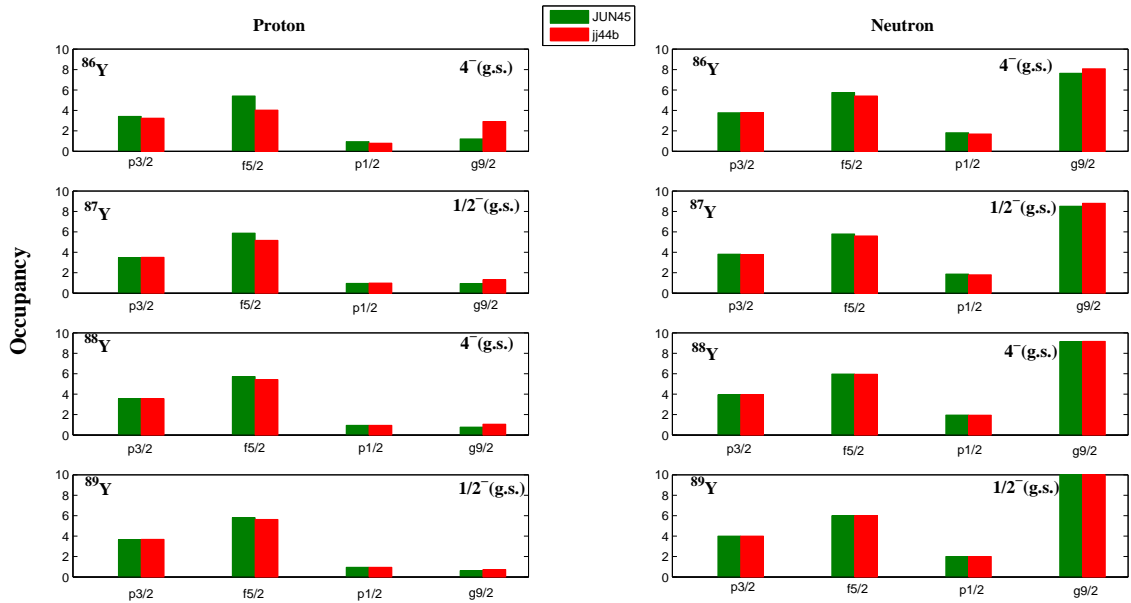} 
}
\caption{Occupancy of proton/neutron orbitals for ground state in $^{86,87,88,89}$Y isotopes.}
\label{fig4}
\end{figure*}

\section{Reduced transition probability, quadrupole and magnetic moments}\label{discuss_AR}

The electric multipoles  of order $L$ are defined as 
\begin{equation}
         B(el,L)= \frac{1}{2J_{i}+1}\mid(J_f\mid\mid \sum_{i} e_{i}r_{i}^L { Y_{L}}(\theta_{i},\phi_{i})\mid\mid\ J_i)\mid^2,
\end{equation}
where 
J$_i$ and J$_f$ are the initial and final state spins, respectively.

The $B(E2)$ values is defined as\\

\begin{equation}
        B(E2)= \frac{1}{2J_{i}+1}\mid(J_f\mid\mid \sum_{i} e_{i}r_{i}^2 { Y_{2}}(\theta_{i},\phi_{i})\mid\mid\ J_i)\mid^2. 
\end{equation} 

The experimentally determined $B(E2)$ values, for different transitions is listed in
table 1. In this table we have also listed value of $E_\gamma$ for corresponding transitions.
The theoretical calculations were performed employing effective charges:
e$_{\rm eff}^\pi$ = 1.5 $e$, e$_{\rm eff}^\nu$ = 0.5 $e$ for two set of effective interactions.
The overall results for JUN45 interaction show better agreement with experimental data.

In table 2, we have also compare quadrupole and magnetic moments with available experimental data.
The results are close to experimental data. Thus we may conclude that the present model space
is sufficient to predict this property. Their is no need to include proton $f_{7/2}$ orbital
in the model space to see the importance of the proton excitation across $Z=28$ shell.

\begin{table}
\begin{center}
\caption{ Electric quadrupole moments, $Q_s$ (in eb), the effective
charges $e_p$=1.5 $e$  , $e_n$=0.5 $e$  are used in the calculation and magnetic moments, $\mu$ (in $\mu_N$), for $g_{s}^{eff}$ = 0.7$g_{s}^{free}$. }
\vspace{2mm}
\label{tab:table3}
\begin{tabular}{ c c c  c  c c } \hline
 & $^{86}$Y &$^{88}$Y&$^{87}$Y &$^{89}$Y& \\ \hline
 & $Q$($8_1^+$)& $Q$($8_1^+$)& $Q$($9/2_1^+$) &  $Q$($9/2_1^+$) &\\ \hline
Experiment & +N/A  & +0.06(6) & -0.50(6) & -0.43(6)  \\ 
JUN45 & -0.16  & -0.022 & -0.44& -0.27 \\ 
jj44b & -0.03  & +0.005 & -0.48 & -0.34\\ 
\hline
& &  & $\mu$($9/2_1^+$) & $\mu$($9/2_1^+$)\\ \hline
 Experiment &   &  & +6.24(2)  & +6.37(4) \\ 
 JUN45 &  &  & +6.59  & +6.81 \\ 
 jj44b &  &  & +6.42  & +6.56 \\ 
       &  &  &        &  \\ \hline
\label{t_q}     
\end{tabular}\end{center}
\end{table}

\section{Occupancy}\label{discuss_occu}

We have plotted occupancy of proton/neutron orbitals for $^{86,87,88,89}$Y in Fig. 5. 
As we move from $^{86}$Y to $^{89}$Y the proton occupancy for $f_{5/2}$ orbital is
increasing while the occupancy of $g_{9/2}$ orbital is decreasing. While in the case of the neutrons, the occupancy of both 
$f_{5/2}$ and $g_{9/2}$ orbitals is increasing. This reflects that as we move towards $^{89}$Y, the neutrons prefer to
occupy $g_{9/2}$ orbital. The occupancy of neutron $p_{1/2}$ orbital is greater than
proton $p_{1/2}$ orbital and it is continuously increasing in the case of jj44b interaction.
The $p_{3/2}$ orbital occupancy is always increasing for both interactions as we
move towards $^{89}$Y. 

\section{Conclusions}\label{fin}
In the present work we performed full-fledged shell model calculation for $^{86,87,88,89}$Y isotopes
in $f_{5/2}pg_{9/2}$ model space. Following conclusions have been drawn from this work:
 
\begin{itemize}

\item{} The calculated energy levels, B(E2)'s, quadrupole and magnetic moments are 
in good agreement with experimental data.

\item{} Further theoretical development for identification of intruder orbitals 
including $d_{5/2}$ orbital in the model space  is needed to study excitation
across $N=50$  shell for better identification of the structure of these nuclei.

\item{} Both interactions predicting  $ \pi (p_{1/2}^1) $ configuration for ground
state ($1/2^-$) for $^{87,89}$Y, while in case of
$^{88}$Y ground state ($4^-$) has $ \pi (p_{1/2}^1) \otimes \nu (g_{9/2}^{-1})$ configuration.

\item{} For the ground state the neutron occupancy is increasing for $f_{5/2}$ and $g_{9/2}$
orbitals as we move from $^{86}$Y to $^{89}$Y.

\end{itemize}
\section*{Acknowledgment}
 All the shell-model calculations have been carried out at TLAPOA computational facility
at ICN, UNAM, Mexico. This work is supported by CSIR fellowship of India.
MJE acknowledges support from the grant No. F2-FA-F177 of 
Uzbekistan Academy of Sciences.

\end{document}